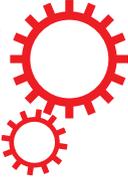

**OPEN**

# Local particle-hole pair excitations by SU(2) symmetry fluctuations

X. Montiel[1,2], T. Kloss[1,3] & C. Pépin[1]

Understanding the pseudo-gap phase which opens in the under-doped regime of cuprate superconductors is one of the most enduring challenges of the physics of these compounds. A depletion in the electronic density of states is observed, which is gapping out part of the Fermi surface, leading to the formation of mysterious lines of massless excitations- the Fermi arcs. Here we give a new theoretical account of the physics of the pseudo-gap phase in terms of the emergence of local patches of particle-hole pairs generated by SU(2) symmetry fluctuations. The proliferation of these local patches accounts naturally for the robustness of the pseudo-gap phase to disturbances like disorder or magnetic field and is shown to gap out part of the Fermi surface, leading to the formation of the Fermi arcs. Most noticeably, we show that these patches induce a modulated charge distribution on the Oxygen atoms, in remarkable agreement with recent X-ray and STM observations.

The concept of symmetries governing the behaviour of physical states is maybe the most robust in theoretical physics. From the formation of nuclei to the Higgs-Boson it has been instrumental in the determination of every emerging state in high energy physics. It would be quite remarkable if a phenomenon as complex as high temperature superconductivity would be governed as well by an overall emergent symmetry. Suggestions about the existence of a pseudo-spin symmetry in the background of the physics of cuprates have been introduced since the early days of these compounds[1] and has been revived over the years in different contexts. In all cases, the main simple idea is that one can rotate the $d$-wave superconducting (SC) state towards another state of matter quasi-degenerate in energy, like anti-ferromagnetism (AF)[2,3], a nematic state[4], or else alternating loop-currents of $d$-density wave[5] or $\pi$-flux phases[6,7]. The physics is then controlled solely by the powerful constraint of the emergent symmetry which produces a vast region of the phase diagram where the fluctuations between those two states are dominant. Here, we argue that an emerging SU(2) symmetry connecting the $d$-wave SC state with a $d$-wave charge order (CO) is the main ingredient of the physics of the under-doped (UD) region, and that the very specific fluctuations associated with this symmetry are responsible for the opening of a gap in the anti-nodal (AN) zone -i.e. $(0, \pi)$ region- of the first Brillouin zone (BZ) leading to the formation of Fermi arcs in the nodal (N) zone i.e. the $(\pi, \pi)$ region initially observed by a depletion in the electronic density of states[8]. A few theoretical proposals for the pseudogap (PG), have led to interesting investigations where the axial CDW is associated with axial pair density wave (PDW) pairing order, a finite momentum superconductivity[9–11], or stabilized by short-range AF fluctuations[12,13]. A state where charge and spin degrees of freedom are separated (fractionalized Fermi liquid) has also been proposed to explain the PG phase[14].

The SU(2) symmetry is realized explicitly in two microscopic models: the $t$-$J$ model at half filling[1] (for AF wave vector) and also the spin-fermion "hotspot" model[15,16] with a linearized electron dispersion. In both models, AF interactions are at the origin of the emergent SU(2) symmetry- which is likely to be the generic case for UD cuprates. For the purpose of this study, the strong-coupling character of the first model, or the vicinity of an AF quantum critical point in the second one are not crucial contrary to the additional symmetry induced in both. Generically, the degeneracy, or quasi-degeneracy in energy of two states can be either accidental or controlled by a symmetry here a pseudo-spin SU(2) symmetry which rotates from a $d$-wave SC state to a $d$-wave CO state (The definition of the SU(2) symmetry is detailed in the SI). The SU(2) fluctuations, typically captured within the O(4) non-linear $\sigma$-model[3,16–18], involve phase fluctuations within each state but also between the two pseudo-spin states. The presence of the SU(2) symmetry in the background of the UD region implies that at some intermediate energy scale, the two pseudo-spin states are indistinguishable. When both, $d$-wave SC and CO pseudo-spin states are quasi degenerate due to thermal fluctuations of the order of the PG temperature $T^*$, the SU(2) pairing fluctuations trigger an instability towards a less symmetric state.

[1]IPhT, L'Orme des Merisiers, CEA-Saclay, 91191, Gif-sur-Yvette, France. [2]Department of Physics, Royal Holloway, University of London, Egham, Surrey, TW20 0EX, United Kingdom. [3]INAC-PHELIQS, Université Grenoble Alpes and CEA, 38000, Grenoble, France. X. Montiel, T. Kloss and C. Pépin contributed equally to this work. Correspondence and requests for materials should be addressed to C.P. (email: catherine.pepin@cea.fr)

## Model

Our starting point is an effective action $S[\psi] = S_0[\psi] + S_1[\psi]$, where the electrons interact with modes subjected to the SU(2) symmetry:

$$S_0[\psi] = -\sum_{k,\sigma} \overline{\psi}_{\sigma\mathbf{k}} G_0^{-1}(k) \psi_{\sigma\mathbf{k}}, \qquad (1)$$

$$S_1[\psi] = \sum_{k,q} [\Delta_{k,q} \overline{\psi}_{\uparrow\mathbf{k}+\mathbf{q}} \overline{\psi}_{\downarrow-\mathbf{k}} + h.c.]. \qquad (2)$$

Here, $\psi$ represents spin $\sigma$ fermions with bare propagator $G_0^{-1}(k) = i\varepsilon_n - \xi_\mathbf{k}$, where $\xi_\mathbf{k}$ represents the dispersion with subtracted chemical potential and $S_1[\psi]$ accounts for the electron-SC interaction mediated through bosonic field $\Delta_{k,q} \sim \langle \psi_{\mathbf{k}+\mathbf{q}/2,\sigma} \psi_{-\mathbf{k}+\mathbf{q}/2,-\sigma} \rangle$. For simplicity we focus on the pairing part of the SU(2) propagator and its effect onto the formation of patches of charge modulations. The reciprocal effect of the charge sector will create a PDW, and will be treated in a more detailed version of this paper.

This bosonic field corresponds to SU(2) pairing fluctuations of a small momentum $q$, with the form typical of the O(4) non linear $\sigma$-model describing the thermal fluctuations between $d$-wave SC and CO states[15–18]:

$$\pi_{k,k',q} = \langle \overline{\Delta}_{k,q} \Delta_{k',q} \rangle = \frac{\overline{\pi}_0}{\omega_n^2 + \overline{J}_1(\mathbf{v}_\mathbf{k} \cdot \mathbf{q})^2 + \overline{a}_{0,\mathbf{k}}}, \qquad (3)$$

where $\overline{\pi}_0, \overline{J}_1, \overline{a}_{0,\mathbf{k}}$ are non-universal parameters and $\mathbf{v}_\mathbf{k}$ the Fermi velocity. The influence of the curvature to the mass is modeled by a contribution to $\overline{a}_0$ (Eqn. (3)) of the form $\sim(\xi_{\mathbf{k}+\mathbf{Q}_0} + \xi_\mathbf{k})^2$, where $\mathbf{Q}_0$ is the diagonal wave vector relating two hot-spots on opposite Fermi surface accross the zone edge[16, 19]. The SU(2) symmetry is realized when $\xi_{\mathbf{k}+\mathbf{Q}_0} = -\xi_\mathbf{k}$, which favours the AN zone. Integrating out the bosonic field $\Delta_{k,q}$ yields a new effective two body electron-electron interaction of the form:

$$S_{\text{fin}}[\psi] = -\sum_{kk'q,\sigma} \pi_{k,k',q} \overline{\psi}_{\sigma,\mathbf{k}} \psi_{\sigma,\mathbf{k}'} \overline{\psi}_{-\sigma,-\mathbf{k}+\mathbf{q}} \psi_{-\sigma,-\mathbf{k}'+\mathbf{q}}. \qquad (4)$$

## Linear response

The form of the effective interaction Eqn. (4) leads to an attractive coupling between the electrons. Its effect on the linear response of the charge susceptibility is visualized in Fig. 1. We observe a drastic increase of the response at zero momentum, accompanied by a continuous series of maxima in a locus of wave vectors following the Fermi surface at $2\mathbf{k}_F$ in the anti-nodal region of the Brillouin Zone (BZ). The zero momentum peak is a signature that the system is on the brink of a nematic transition, whereas the continuum of peaks at momentum $2\mathbf{k}_F$ signals a *new* kind of instability of the particle-hole response.

## Excitonic pairing mechanism

To get a simple insight, let consider the electron-hole wavefunction $\chi_{\mathbf{r},\mathbf{r}'}$:

$$\chi_{\mathbf{r},\mathbf{r}'} = \sum_{\mathbf{P},\mathbf{k},\sigma} e^{\left[-i\mathbf{P}\cdot\frac{\mathbf{r}+\mathbf{r}'}{2}\right]} e^{[i\widetilde{\mathbf{k}}\cdot(\mathbf{r}-\mathbf{r}')]} \chi_{\widetilde{\mathbf{k}},\mathbf{P}}, \qquad (5)$$

with $\widetilde{\mathbf{k}} = 2\mathbf{k} - \mathbf{P}$, $\mathbf{P}$ the modulation wave vector at $\mathbf{P} = \{2\mathbf{k}_F\}$ which connects two opposite sides of the Fermi surface and $\chi_{\widetilde{\mathbf{k}},\mathbf{P}}$ assumed as a constant. Inserting the electron-hole wave function $\chi_{\mathbf{r},\mathbf{r}'}$ (equation (5)) into the two particles Schrödinger equation, we find

$$\left[-\frac{\hbar^2}{2m}(\partial_\mathbf{r}^2 - \partial_{\mathbf{r}'}^2) + V_{\mathbf{r},\mathbf{r}'}\right] \chi_{\mathbf{r},\mathbf{r}'} = E \chi_{\mathbf{r},\mathbf{r}'}, \qquad (6)$$

that leads to $\left(E - \frac{\hbar^2}{m} \mathbf{k}_F \cdot \widetilde{\mathbf{k}}\right) \chi_{\widetilde{\mathbf{k}},\mathbf{P}} = C$, with $C = \sum_{\widetilde{\mathbf{k}}} V_{\widetilde{\mathbf{k}},\widetilde{\mathbf{k}}'} \chi_{\widetilde{\mathbf{k}},\mathbf{P}}$, where $V_{\widetilde{\mathbf{k}},\widetilde{\mathbf{k}}'} = \int_\mathbf{r} V(r) e^{i(\widetilde{\mathbf{k}}-\widetilde{\mathbf{k}}')\cdot\mathbf{r}}$ is an attractive potential coming from the pairing fluctuations. With $\omega_F$ the width of the fluctuation spectrum, we model $V_{\widetilde{\mathbf{k}},\widetilde{\mathbf{k}}'} = -V$ if both momenta $0 < \widetilde{\mathbf{k}} < \omega_F/\rho_0$ and $0 < \widetilde{\mathbf{k}}' < \omega_F/\rho_0$ ($\rho_0$ being the electronic density of states) and $V_{\widetilde{\mathbf{k}},\widetilde{\mathbf{k}}'} = 0$ elsewhere. Eqn. (6) can then easily be solved, leading to the bonding energy

$$E = -2\hbar\omega_F e^{-2/(\rho_0 V)}. \qquad (7)$$

The formation of particle-hole pairs at multiple $2\mathbf{k}_F$ wave vectors is a logarithmic instability of the Fermi liquid in the presence of an attractive potential. In the standard BCS theory, the coupling between density and phase fluctuations is weak. In some specific cases, however, like the attractive Hubbard model, density and phase couple strongly and our model likewise predicts the emergence of s-wave particle-hole pair patches (see the SI for more details). Within the SU(2) scenario for Cuprates, the typical scale associated the pairing fluctuations is strong, of order of the formation of the SU(2) dome, and can naturally be associated with the PG scale $T^*$.

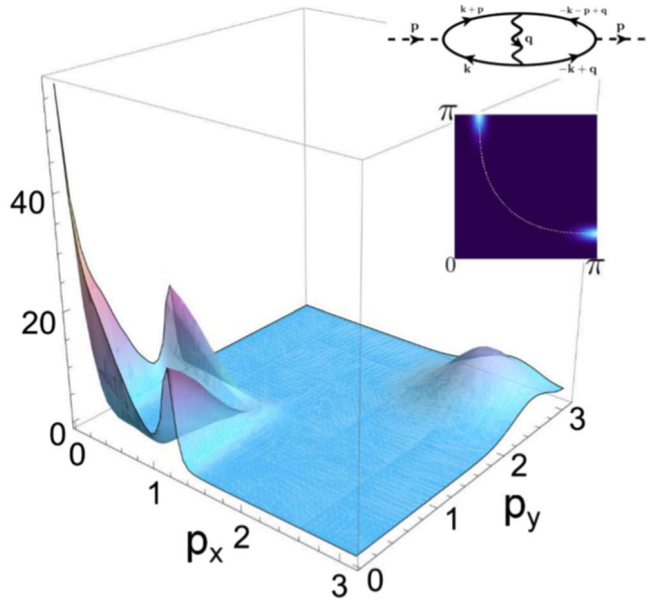

**Figure 1.** Real part of the charge susceptibility in presence of SU(2) pairing fluctuations. The charge susceptibility is maximal at $p=0$ which is the signature of a nematic instability and is non zero along the $2\mathbf{p}_F$ vectors. Note that it exhibits local maxima along $p_x$ and $p_y$ axes that correspond to the $2\mathbf{p}_F$ vectors of the zone edge. Inset: we draw the Feynam diagram of the charge susceptibility and the mass $\bar{a}_0$ in the positive quarter of the first BZ (the dashed line represents the Fermi surface). The real part of the charge susceptibility is determined after integration over internal Matsubara frequencies at zero temperature and the summation over internal momentum has been performed numerically in the first Brillouin zone with meshes of $200 \times 200$ at zero frequency. Regarding the bosonic propagator written in equation (3), the main contribution to charge susceptibility is raised for $\mathbf{q} \approx 0$.

## Solution of the gap equation

Eqn. (4) can now be decoupled in the charge channel. Upon introducing the collective field

$$\chi_{k,k'} = T \sum_{q,\omega_n} \pi_{k,k',q} \langle \overline{\psi}_{\uparrow -\mathbf{k}+\mathbf{q}} \psi_{\uparrow -\mathbf{k}'+\mathbf{q}} \rangle, \tag{8}$$

representing a generalized charge order at wave vectors $\mathbf{k}$ and $\mathbf{k}'$, the self-consistent mean-field equation yields

$$\chi_{k,k'} = -T \sum_{q,\omega_n} \frac{\pi_{k,k'q} \chi_{q-k,q-k'}}{(i\omega_n - \xi_{\mathbf{q}-\mathbf{k}})(i\omega_n - \xi_{\mathbf{q}-\mathbf{k}'}) - |\chi_{q-k,q-k'}|^2}. \tag{9}$$

Eq. (9) is solved numerically to find the stability region for the particle-hole pair at wave vectors $\mathbf{k}$ and $\mathbf{k}'$, and depicted in Fig. 2.

The $(\mathbf{k}, \mathbf{k}')$-points for which $\chi_{\mathbf{k},\mathbf{k}'}$ is non zero have the peculiarity that the outgoing wave vector $\mathbf{k}'$ is related to the ongoing wave vector $\mathbf{k}$ through the relation $\mathbf{k}' - \mathbf{k} = 2\mathbf{p}_F$, which leads to the formation of a *local* particle-hole pair mode-called Resonant Excitonic State (RES), analogous to a soliton. The fact that the preferred wave vectors $2\mathbf{p}_F$ are peaked on the Fermi surface comes from the denominator of Eqn. (9), which is minimal for those wave vectors. Note the non-linearity in the field $\chi$ in Eqn. (9), which is a crucial ingredient for stabilizing the local solution. The situation is similar to electrons interacting with a wide range of bosonic modes leading to the formation of local polarons[20], with particle-hole pairs instead of electrons. The bosons are the local pairing fluctuations of the non-linear $\sigma$-model Eqn. (3), as opposed to the polaronic scenario, where electrons are coupled to phonons with wide spectrum. The SU(2) symmetry is the mechanism controlling the coupling between pairing and density fluctuations, which typically leads to the formation of non-linear equations of motion and the emergence of local modes (see e.g. ref. [21]).

The momentum dependence of the order parameter $\chi_{\widetilde{\mathbf{k}},\mathbf{P}} = \chi_0^{\mathbf{P}} F_{\mathbf{k}} \sim \langle \overline{\psi}_{\mathbf{k},\sigma} \psi_{\mathbf{k}-\mathbf{P},\sigma} \rangle$ is fully determined by the equation (9). The form factor $F_{\mathbf{k}}$ measures the phase space allocated to each $\mathbf{P}$ modulation, with $|\widetilde{\mathbf{k}}| \leq |F_{\mathbf{k}}|$, as shown in Fig. 2b). The natural *local* structure of the particle-hole field $\chi_{\mathbf{k},\mathbf{k}'}$ is due to the summation on the multiple P-vectors, which forces the center of mass of the particle-hole pair to rest at $x=0$ and makes it similar to a soliton solution of the nonlinear Eq. (9). This local object, or patch-, then has the tendency to proliferate, and while interacting with its surroundings will induce phase separation. From another viewpoint, the non-linear sigma-model, which describes non-linear interactions between the charge and superconducting sectors, produces such local modes in the form of topological defects. The relationship between the particle-hole patches and topological defects will be studied in a further work. Translation symmetry is broken, but in contrast to the standard charge ordering, it is broken *locally*, which does not imply global periodicity.

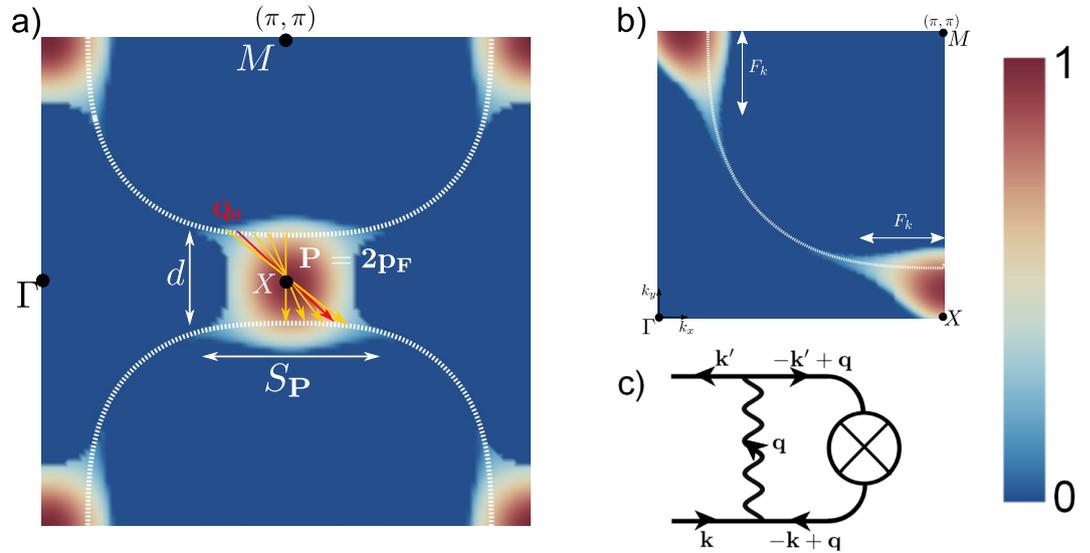

**Figure 2.** Amplitude of the particle-hole pair patch order parameter $|\chi_{\mathbf{k},\mathbf{k}'}|$ in the first Brillouin zone. (**a**) Color plot of $|\chi_{\mathbf{k},\mathbf{k}'}|$ in the AN zone $(\pi, 0)$, as a function of the ordering wave vector $\mathbf{q}_c (\mathbf{k}' = \mathbf{k} - \mathbf{q}_c)$ (normalized to the maximum value). The amplitude of the electron-hole pair is maximum in the AN region selecting a set of degenerate solutions at $2\mathbf{p}_F$ wave vectors. It is suppressed in the nodal region $(\pi, \pi)$. We represent the quasi-degenerate $2\mathbf{p}_F$ wave vectors of the electron-hole pairs on opposed Fermi surfaces in the AN region, that give rise to the RES state. The vector $Q_0$ is the diagonal ordering vector (red vector). The momentum spread is $S_\mathbf{p}$. (**b**) Density plot of the particle-hole pair amplitude $|\chi_{\mathbf{k},\mathbf{k}'}|$ for $\mathbf{k}' = \mathbf{k} + 2\mathbf{p}_F(\mathbf{k})$ (normalized to the maximum value). The size of the gapped zone depends on the magnitude of the interaction $\pi_0$ as demonstrated in the Figure S2. Here, the figure are presented for $\overline{\pi}_0 = 1$, the other parameters are exposed in the SI. (**c**) Feynman diagram representing the $\chi_{\mathbf{k},\mathbf{k}'}$ mean field equation (9).

Remarkably, each patch resembles very much a superconductor, but made of electron-hole pairs instead of Cooper pairs (electron-electron pairs). The typical size of the pair - also called "coherence length" in the SC analogy, is typically small with $\xi \sim \hbar v_F/(\pi \chi_{AN})$ (with $\chi_{AN}$ the typical amplitude of the AN gap), which leads to the preferential formation of pair on nearest neighbor Cu-bonds. The second typical length corresponds to the size of the patches, which we call $L_\mathbf{p}$. It is controlled by the spread $S_\mathbf{p}$ of the **P** wave vectors as shown in Fig. 2b). The typical patch size of the RES $L_\mathbf{p} \sim 2\pi \hbar / S_\mathbf{p}$ is of order of a few lattice sites.

## Implications for an STM experiment

A recent set of experiments has shed a new light on the mysterious PG phase of the Cuprates. Charge modulations with incommensurate wave vectors parallel to the axes $\mathbf{Q}_x = \pm \frac{2\pi}{a_0}(0, d)$, $\mathbf{Q}_y = \pm \frac{2\pi}{a_0}(d, 0)$ with $d \sim 0.3$ have been observed in this doping range through various experimental probes[22–36] (with $a_0$ the elementary cell parameter of the square lattice). A priori there is no obvious relation between the incipient charge order and the PG phase of the Cuprates. However, a recent set of data obtained using the Scanning Tunnelling Microscope (STM) argue otherwise[37]. A careful study of the form factor associated to the charge signal has allowed to distinguish the s-wave part from the d-wave part of the charge modulation located on the Oxygen atoms, both from STM, NMR and X-ray experiments[32, 36–38]. The signal, shows a very striking behaviour as a function of the applied voltage. Its s-wave part (10% of the signal at $T^*$) shows a typical energy scale of the order of the superconducting gap. Its d-wave part (90%), however, is strongly peaked at the PG energy scale. This result links in a very unexpected way the physics of the PG with the modulations observed in the charge sector, as if they were coherent only below a certain energy scale, but their pure d-wave part persists up to the PG scale in a very incoherent manner, detectable only by local probes like STM.

We present here a simple explanation for this remarkable observation. We evaluate the charge density on the oxygen O(2p) orbitals of the the $CuO_2$ lattice induced by the RES, which writes

$$\rho_{O,i} = \frac{1}{N}\sum_{\mathbf{k},\mathbf{P}}\left[\cos\left(\left(\mathbf{k} + \frac{\mathbf{P}}{2}\right)\mathbf{a}_0^i\right)\cos(\mathbf{P}.\mathbf{R})\right.$$
$$\left. \times \left(\left\langle \psi_{\mathbf{k}+\frac{\mathbf{P}}{2},\sigma}^\dagger \psi_{\mathbf{k}-\frac{\mathbf{P}}{2},\sigma}\right\rangle + h.c\right)\right] \quad (10)$$

with $i = \{x, y\}$, $N$ the number of copper sites and the correlator $\left(\left\langle \psi_{\mathbf{k}+\frac{\mathbf{P}}{2},\sigma}^\dagger \psi_{\mathbf{k}-\frac{\mathbf{P}}{2},\sigma}\right\rangle + h.c\right)$ is related to the RES Green function. We first focus on the local symmetries around the copper sites, with for d-wave $\rho_O^d = (\rho_{Ox} - \rho_{Oy})/2$ and for s'-wave $\rho_O^{s'} = (\rho_{Ox} + \rho_{Oy})/2$ components. The spatial dependence of the charge den-sity on O orbitals is presented in Fig. 3(a,b) where one can notice the checkerboard structure with preponderant axial wave vectors $\mathbf{Q_x}$ and $\mathbf{Q_y}$. The signal is found to be 98% $d$-wave, in good agreement with the experimental data. The $d$-wave form factor originates the diagonal part of the charge order modulation while the s′-wave form factor accounts for non-diagonal modulations. Note that a pure diagonal charge order would produce a pure $d$-wave form factor[16]. We then Fourier transform the $d$-wave component and report the result in Fig. 4. Rather large spots centered on $|\mathbf{Q_x}| = |\mathbf{Q_y}| = 0.237\frac{2\pi}{a_0}$ are found, which agree in a remarkable way with the observation of ref. 37 that the pure $d$-wave incoherent part of the axial charge modulation extends up to the PG energy. This internal modulation of each patch comes from the summation over the ordering vectors **P** in equation (10). As shown in Fig. 2(a), the modulation spreading is on the y-direction for the (0, π)- region and respectively in the x-direction in the $(\pi, 0)$- region. This leads to the typical checkerboard form, with two average modulation wave vectors around $\mathbf{Q}_x = \pm 2\pi/a_0(d, 0)$ and $\mathbf{Q}_y = \pm 2\pi/a_0(0, d)$ with $d \sim 0.3$ (see Fig. 2(b)), which are expected to be visible through local probes like STM. The key feature in our model is that diagonal wave vectors are much disfavoured compared to parallel ones, because they have very small phase space in the extension zone $F_k$ where the RES order parameter $\chi^{RES}$ is non zero (see Fig. 2). Hence very remarkably, our proposal for the particle-hole pair state has a preferential structure in k-space dominated by incoherent wave vectors centered on $\mathbf{Q}_x$ and $\mathbf{Q}_y$ rather than on the diagonal. We are aware that the STM experiments show patches with strong x/y anisotropy accompanied with a C4 symmetry breaking. We have not focused here on this aspect of the experiment, preferring to explain how to use the underlying SU(2) symmetry to account for the emergence of charge modulations at the PG energy scale. The theory is easily generalized to the uniaxial case, with a set of SU(2) operators (defined in the SI) for each wave vector $Q_x$ or $Q_y$.

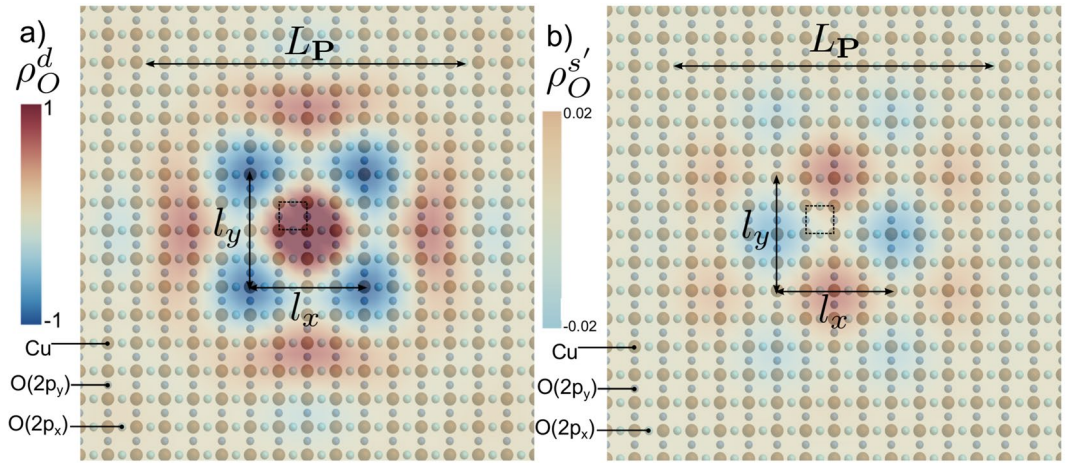

**Figure 3.** Charge density on O atoms. (**a**) Visualization of the $d$-wave component of the charge density on oxygen atoms $\rho_O^d$ normalized to $\pm 1$ value in the real space. (**b**) Visualization of the s′-wave component of charge density on oxygen atoms $\rho_O^{s'}$, normalized to the maximal value of $\rho_O^d$ in the real space. In both $\rho_O^d$ and $\rho_O^{s'}$, we see an incommensurate charge order developing on oxygen atoms with wave length $l_x = 2\pi/Q_x$ and $l_y = 2\pi/Q_y$. The elementary cell of the square lattice is represented by the dashed square. The Fig. 3(a,b) have been determined by evaluating the formula (10). The summation over the momentum **k** has been performed in the FBZ while the summation over **P** vectors has been done over two-hundred ordering vectors **P**. We choose a distribution of vectors **P** around the Fermi surface. The extension in momentum of these distribution is the one presented in Fig. 2(a). The calculation has been done considering the electronic dispersion for Bi2212 compounds used in refs 33, 37 at hole doping $p = 0.08$. The coherence length of the particle hole pair is of order of the length between two Copper site $\xi \approx a_0$ while the size of the patch is larger, $L_P \approx 10 a_0$.

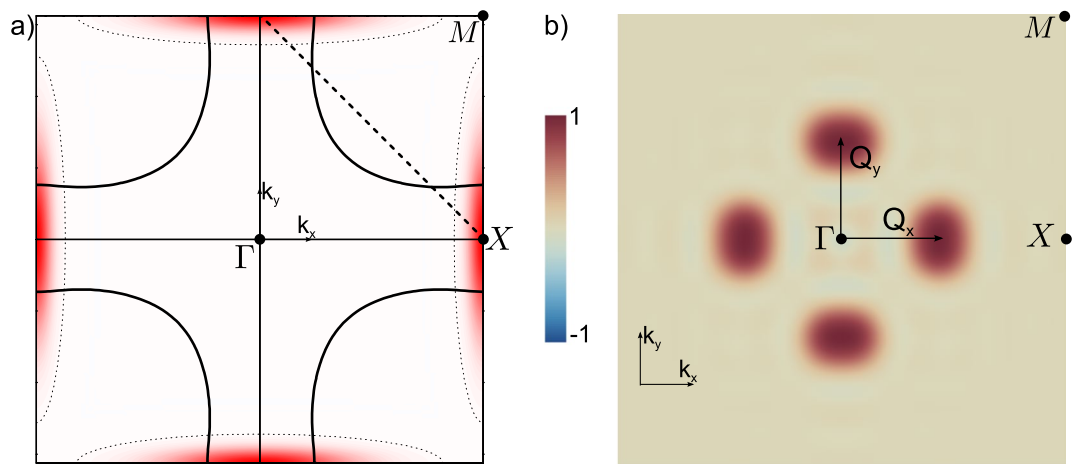

**Figure 4.** Spectral analysis of $\rho_O^d$. Presentation of the spectral analysis of the $d$-wave component of the charge density on oxygen atoms. In (**a**), we present the magnitude of gap envelop $F_k$ in the first Brillouin zone of the square lattice (red area delimited by the dotted line). (**b**) Fourier transform of the charge density of Fig. 3(a) in the first Brillouin Zone. The charge order is organized preferentially along $Q_x$ and $Q_y$ that are parallel to $k_x$ and $k_y$ respectively.

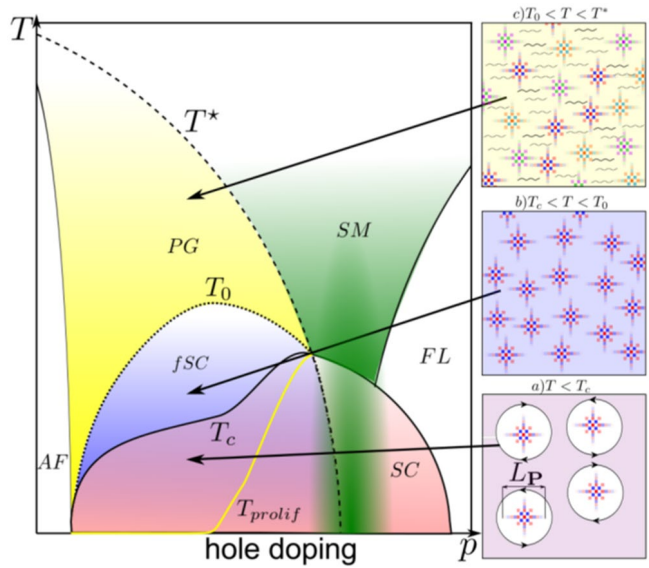

**Figure 5.** Schematic temperature-hole doping (T, P) phase diagram of cuprate superconductors. The strange metal phase is depicted in green and the Fermi liquid (FL) in white. *Insets*: (**a**) For $T < T_c$, the particle-hole pair patches co-exist with superconductivity and are especially visible inside the vortex-antivortex pairs. $L_P$ is the typical size of a particle-hole pair patch. (**b**) For $T_c < T < T_0$, one observes the proliferation of patches and the superconducting fluctuations (fSC)- fluctuating up to a U(1) subset of the SU(2) symmetry- disappear at $T_0$. (**c**) For $T_0 < T < T^*$, the patches modulations are incoherent and electron scattering through the "particle-hole pair soup" leads to dissipation which accounts for the increase of the Fermi arcs with $T$. The temperature of proliferation of the patches $T_{prolif}$ is presented in yellow.

### The global phase diagram of high-Tc superconductors

We propose a real space picture which describes how the proliferation of patches is responsible for the opening of the PG (see Fig. 5). First, the proliferation of local objects like particle-hole pair patches leads to a state very robust to perturbations by strong disorder[8] or magnetic field[30, 36, 39, 40], which puts strong restrictions on theoretical proposals. Then, our theory belongs to the class of "one-gap" scenarios, for which the coherent SC state is destroyed at $T_c$ by fluctuations[41–43], but the class of fluctuations at play here have SU(2)-character, as opposed to only the U(1)-type preformed pair scenario. Proliferation of particle-hole pair patches destabilizes the AN part of the Fermi surface, leading to the formation of Fermi arcs whose lengths is increasing with $T$ up to $T^*$, as reported in Angle Resolved Photo Emission (ARPES) (see e. g. refs 43–45).

The temperature at which the patches start to proliferate is proportional to the difference between the binding energies of particle-hole pairs and Cooper pairs $T_{prolif} \sim Max[E_P - E_C, 0]$ (see Fig. 5). SU(2) pairing fluctuations vanish around optimal doping, leading to a maximal $T_{prolif} \sim T_c$, in contrast to heavily underdoped compounds ($x < 0.12$) where, in the vicinity of the Mott localization, particle-hole pairing beats Cooper pairing ($E_P - E_C < 0$), and the proliferation of patches occurs at almost zero temperature. In this regime the whole anti-nodal region of the BZ is gapped out by the particle-hole pair patches, leading to a fractionalization of the Fermi surface into two gaps.

A typical evolution with temperature is as follows. Below $T_c$ ($0 < T < T_c$) as depicted in Fig. 5(a), phase fluctuations destroy the coherent SC state. These fluctuations manifest themselves as local defects of the SC density, that are similar to vortex-antivortex pairs in classical superconductors, but with the difference that in our case, the normal phase is made of local particle-hole pair patches which can be seen inside the vortices. An important point is the one of global phase coherence. Below $T_c$ the global phase of modulations inside each particle-hole pair patch can get locked with the SC one, so that the intrinsic checkerboard modulation of the patches acquires a global phase coherence. This phenomenon enhances the coherent charge density wave (CDW)-like signal experimentally observed at zero magnetic field, and has been recently reported in another STM study[46].

At $T = T_c$, the gap around the node disappears, causing the loss of SC coherence, and at an intermediate temperature $T_0$, such that $T_c < T_0 < T^*$ the standard SC fluctuations are lost, as observed for example, by Josephson effect[47], Nernst effect[48] or by study of the resistivity[49]. The transition towards the SC state can thus be described by a phase correlation length $\widetilde{\lambda}_c$, which diverges at $T_c$ and becomes very small at $T^*$. $\widetilde{\lambda}_c^{-1}$ can be understood as the typical scale of all the phase fluctuations in the system (in the present SU(2) scenario there are three types of phase fluctuations: SC, CO, and the angle between these two).

For $T_c < T < T_0$ (Fig. 5(b)), the global phase coherence of the Cooper pairs and modulated particle-hole pair patches is lost. The SC superfluid density $n_s$ vanishes at $T_0$. In the regime, $T_0 < T < T^*$ shown in Fig. 5(c), the RES manifests itself as an incoherent set of particle-hole pair patches. Electronic scattering through the RES induces a finite lifetime, which accounts for the size of the Fermi arcs increasing with $T$[43]. Above $T^*$, the RES disappears.

## Concluding remarks

We propose a new mechanism for gapping out the AN region of the Fermi surface, leading to the formation of Fermi arcs below $T^*$. In our model, an SU(2) symmetry governs the physics of the UD region of the phase diagram, and SU(2) pairing fluctuations allow for the emergence of new local excitations in the form of particle-hole pair patches which possess intrinsic checkerboard modulations. The proliferation of these patches opens a gap in the anti-nodal region of the Brillouin zone, leading to the formation of Fermi arcs. Very importantly, it is accompanied (see Fig. 1) with the formation of a $q = 0$ order in the nematic sector, intensely discussed at the moment[50]. The polarizable structure of the patches can also lead to the formation of intra unit cell orders[51, 52]. One striking feature of our theory, is that it reconciles the one-gap vs. two gap scenarios: fluctuations destroy the SC phase at $T_c$, creating Fermi arcs, as in the one-gap scenario[7, 43], but on the other hand the AN and nodal regions behave very differently upon raising the temperature (with particle-hole pair patches forming in the AN region), which is reminiscent of two-gaps. Note that typical two gaps scenarios produce Fermi pockets instead of Fermi arcs in the nodal region[53, 54]. The SU(2) symmetry is identically verified in a very simplified model where the Fermi surface is reduced to eight points or "hot spots" and where the dispersion of the electrons has been linearized[15, 16], and the electrons interact through long range anti-ferromagnetic (AF) modes within the spin-fermion model[55]. The SU(2) paradigm, though, is more robust than AF quantum critical point -or a zero temperature phase transition with long range fluctuations of the AF modes, in the sense that the symmetry can be in the background of the under-doped region even though the cuprates show only short range AF modes. Recent pump-probe experiments exhibiting phase fluctuating SC/CDW composite order, give support to this idea[56]. The importance of anti-ferromagnetism though, should not be underestimated, as it supports the SU(2) pairing fluctuations predominantly in the anti-nodal region of the BZ[57] and originates the $d$-wave nature of both SC and charge order states. The recent findings that the modulation **q**-space are intrinsically related to the formation of the PG[37, 45] are one of the most unexpected developments in the history of the physics in cuprates. The new particle-hole pair state -RES- proposed here possesses intrinsically this feature, and this can only let us wonder whether such a particle-hole pair state could be the solution to the mystery of the PG. The absence of electron pockets in the AN zone of the first BZ in the overdoped part of the phase diagram could provide a check of our scenario and could be performed by ARPES. Moreover, one possible specific signature of such new state could be the presence of photoluminescence signal.

## References

1. Yang, C. N. $\eta$ pairing and off-diagonal long-range order in a hubbard model. *Phys. Rev. Lett.* **63**, 2144–2147, doi:10.1103/PhysRevLett.63.2144 (1989).
2. Demler, E. & Zhang, S.-C. Theory of the resonant neutron scattering of high-$T_c$ superconductors. *Phys. Rev. Lett.* **75**, 4126–4129, doi:10.1103/PhysRevLett.75.4126 (1995).
3. Zhang, S. C. A unified theory based on SO (5) symmetry of superconductivity and antiferromagnetism. *Science* **275**, 1089–1096, doi:10.1126/science.275.5303.1089 (1997).
4. Kee, H.-Y., Doh, H. & Grzesiak, T. Intimate relations between electronic nematic, $d$-density wave and $d$-wave superconducting states. *J. Phys. Condens. Matter* **20**, 255248, doi:10.1103/PhysRevLett.97.257001 (2008).
5. Nayak, C. $O(4)$-invariant formulation of the nodal liquid. *Phys. Rev. B* **62**, R6135–R6138, doi:10.1103/PhysRevB.62.R6135 (2000).
6. Kotliar, G. & Liu, J. Superexchange mechanism and $d$-wave superconductivity. *Phys. Rev. B* **38**, 5142–5145, doi:10.1103/PhysRevB.38.5142 (1988).
7. Lee, P. A., Nagaosa, N. & Wen, X.-G. Doping a Mott insulator: Physics of high-temperature superconductivity. *Rev. Mod. Phys.* **78**, 17–85, doi:10.1103/RevModPhys.78.17 (2006).
8. Alloul, H., Mendels, P., Casalta, H., Marucco, J. F. & Arabski, J. Correlations between magnetic and superconducting properties of Zn-substituted YBa$_2$Cu$_3$O$_{6+x}$. *Phys. Rev. Lett.* **67**, 3140–3143, doi:10.1103/PhysRevLett.67.3140 (1991).
9. Lee, P. A. Amperean pairing and the pseudogap phase of cuprate superconductors. *Phys. Rev. X* **4**, 031017, doi:10.1103/PhysRevX.4.031017 (2014).
10. Wang, Y., Agterberg, D. F. & Chubukov, A. Interplay between pair- and charge-density-wave orders in underdoped cuprates. *Phys. Rev. B* **91**, 115103, doi:10.1103/PhysRevB.91.115103 (2015).
11. Wang, Y., Agterberg, D. F. & Chubukov, A. Coexistence of charge-density-wave and pair-density-wave orders in underdoped cuprates. *Phys. Rev. Lett.* **114**, 197001, doi:10.1103/PhysRevLett.114.197001 (2015).
12. Atkinson, W. A., Kampf, A. P. & Bulut, S. Charge order in the pseudogap phase of cuprate superconductors. *New Journal of Physics* **17**, 013025, doi:10.1088/1367-2630/17/1/013025 (2015).
13. Chowdhury, D. & Sachdev, S. Feedback of superconducting fluctuations on charge order in the underdoped cuprates. *Phys. Rev. B* **90**, 134516, doi:10.1103/PhysRevB.90.134516 (2014).
14. Chatterjee, S. & Sachdev, S. Fractionalized fermi liquid with bosonic chargons as a candidate for the pseudogap metal. *Phys. Rev. B* **94**, 205117, doi:10.1103/PhysRevB.94.205117 (2016).
15. Metlitski, M. A. & Sachdev, S. Quantum phase transitions of metals in two spatial dimensions. II. *Spin density wave order. Phys. Rev. B* **82**, 075128, doi:10.1103/PhysRevB.82.075128 (2010).
16. Efetov, K. B., Meier, H. & Pépin, C. Pseudogap state near a quantum critical point. *Nat. Phys* **9**, 442–446, doi:10.1038/nphys2641 (2013).
17. Meier, H., Einenkel, M., Pépin, C. & Efetov, K. B. Effect of magnetic field on the competition between superconductivity and charge order below the pseudogap state. *Phys. Rev. B* **88**, 020506, doi:10.1103/PhysRevB.88.020506 (2013).

18. Hayward, L. E., Hawthorn, D. G., Melko, R. G. & Sachdev, S. Angular Fluctuations of a Multicomponent Order Describe the Pseudogap of $YBa_2Cu_3O_{6+x}$. *Science* **343**, 1336–1339, doi:10.1126/science.1246310 (2014).
19. Kloss, T., Montiel, X. & Pépin, C. SU(2) symmetry in a realistic spin-fermion model for cuprate superconductors. *Phys. Rev. B* **91**, 205124, doi:10.1103/PhysRevB.91.205124 (2015).
20. Ciuchi, S. & de Pasquale, F. Charge-ordered state from weak to strong coupling. *Physical Review B* **59**, 5431–5440, doi:10.1103/PhysRevB.59.5431 (1999).
21. De Palo, S., Castellani, C., Di Castro, C. & Chakraverty, B. K. Effective action for superconductors and BCS-Bose crossover. *Physical Review B* **60**, 564–573, doi:10.1103/PhysRevB.60.564 (1999).
22. Hoffman, J. E. *et al.* A Four Unit Cell Periodic Pattern of Quasi-Particle States Surrounding Vortex Cores in $Bi_2Sr_2CaCu_2O_{8+\delta}$. *Science* **295**, 466–469, doi:10.1126/science.1066974 (2002).
23. McElroy, K. *et al.* Coincidence of Checkerboard Charge Order and Antinodal State Decoherence in Strongly Underdoped Superconducting $Bi_2Sr_2CaCu_2O_{8+\delta}$. *Phys. Rev. Lett.* **94**, 197005, doi:10.1103/PhysRevLett.94.197005 (2005).
24. Doiron-Leyraud, N. *et al.* Quantum oscillations and the Fermi surface in an underdoped high-$T_c$ superconductor. *Nature* **447**, 565–568, doi:10.1038/nature05872 (2007).
25. Wise, W. D. *et al.* Charge-density-wave origin of cuprate checkerboard visualized by scanning tunnelling microscopy. *Nat. Phys.* **4**, 696–699, doi:10.1038/nphys1021 (2008).
26. Harrison, N. & Sebastian, S. E. Protected nodal electron pocket from multiple-**Q** ordering in underdoped high temperature superconductors. *Phys. Rev. Lett.* **106**, 226402, doi:10.1103/PhysRevLett.106.226402 (2011).
27. Ghiringhelli, G. *et al.* Long-Range Incommensurate Charge Fluctuations in $(Y,Nd)Ba_2Cu_3O_{2+x}$. *Science* **337**, 821–825, doi:10.1126/science.1223532 (2012).
28. Blackburn, E. *et al.* X-Ray Diffraction Observations of a Charge-Density-Wave Order in Superconducting Ortho-II $YBa_2Cu_3O_{6.54}$ Single Crystals in Zero Magnetic Field. *Phys. Rev. Lett.* **110**, 137004, doi:10.1103/PhysRevLett.110.137004 (2013).
29. Chang, J. *et al.* Direct observation of competition between superconductivity and charge density wave order in $YBa_2Cu_3O_{6.67}$. *Nat Phys* **8**, 871–876, doi:10.1038/nphys2456 (2012).
30. LeBoeuf, D. *et al.* Thermodynamic phase diagram of static charge order in underdoped $YBa_2Cu_3O_y$. *Nat. Phys.* **9**, 79–83, doi:10.1038/nphys2502 (2013).
31. Comin, R. *et al.* Charge Order Driven by Fermi-Arc Instability in $Bi_2Sr_{2-x}La_xCuO_3$. *Science* **343**, 390–392, doi:10.1126/science.1242996 (2014).
32. da Silva Neto, E. H. *et al.* Ubiquitous Interplay Between Charge Ordering and High-Temperature Superconductivity in Cuprates. *Science* **343**, 393–396, doi:10.1126/science.1243479 (2014).
33. Fujita, K. *et al.* Simultaneous Transitions in Cuprate Momentum-Space Topology and Electronic Symmetry Breaking. *Science* **344**, 612–616, doi:10.1126/science.1248783 (2014).
34. Blanco-Canosa, S. *et al.* Resonant X-ray scattering study of charge-density wave correlations in $YBa_2Cu_3O_{6+x}$. *Phys. Rev. B* **90**, 054513, doi:10.1103/PhysRevB.90.054513 (2014).
35. Tabis, W. *et al.* Charge order and its connection with Fermi-liquid charge transport in a pristine high-$T_c$ cuprate. *Nat. Commun.* **5**, 5875, doi:10.1038/ncomms6875 (2014).
36. Wu, T., Mayaffre, H., Krämer, S. & Horvatić, M. Incipient charge order observed by NMR in the normal state of $YBa_2Cu_3O_y$. *Nature* **6**, 6438, doi:10.1038/ncomms7438 (2015).
37. Hamidian, M. H. *et al.* Atomic-scale electronic structure of the cuprate d-symmetry form factor density wave state. *Nat. Phys* **12**, 150–156, doi:10.1038/nphys3519 (2015).
38. Comin, R. *et al.* Symmetry of charge order in cuprates. *Nature Materials* **14**, 796–800, doi:10.1038/nmat4295 (2015).
39. Chang, J. *et al.* Magnetic field controlled charge density wave coupling in underdoped $YBa_2Cu_3O_{6+x}$. *Nat Commun* **7**, 11494, doi:10.1038/ncomms11494 (2016).
40. Wu, T. *et al.* Emergence of charge order from the vortex state of a high-temperature superconductor. *Nat. Commun.* **4**, 2113, doi:10.1038/ncomms3113 (2013).
41. Ferraz, A. & Kochetov, E. Effective spin-fermion model for strongly correlated electrons. *Europhys. Lett.* **109**, 37003, doi:10.1209/0295-5075/109/37003 (2015).
42. Emery, V. J. & Kivelson, S. A. Importance of Phase Fluctuations in Superconductors with Small Superfluid Density. *Nature* **374**, 434–437, doi:10.1038/374434a0 (1995).
43. Norman, M. R., Randeria, M., Ding, H. & Campuzano, J. C. Phenomenology of the low-energy spectral function in high-$T_c$ superconductors. *Phys. Rev. B* **57**, R11093–R11096, doi:10.1103/PhysRevB.57.R11093 (1998).
44. Shen, K. M. Nodal Quasiparticles and Antinodal Charge Ordering in $Ca_{2-x}Na_xCuO_2Cl_2$. *Science* **307**, 901–904, doi:10.1126/science.1103627 (2005).
45. He, R.-H. *et al.* From a single-band metal to a high-temperature superconductor via two thermal phase transitions. *Science* **331**, 1579–1583, doi:10.1126/science.1198415 (2011).
46. Hamidian, M. H. *et al.* Magnetic-field Induced Interconversion of Cooper Pairs and Density Wave States within Cuprate Composite Order. *ArXiv:1508.00620* (2015).
47. Bergeal, N. *et al.* Pairing fluctuations in the pseudogap state of copper-oxide superconductors probed by the Josephson effect. *Nat. Phys* **4**, 608–611, doi:10.1038/nphys1017 (2008).
48. Wang, Y., Li, L. & Ong, N. P. Nernst effect in high-$T_c$ superconductors. *Phys. Rev. B* **73**, 024510, doi:10.1103/PhysRevB.73.024510 (2006).
49. Rullier-Albenque, F., Alloul, H. & Rikken, G. High-field studies of superconducting fluctuations in high-$T_c$ cuprates: Evidence for a small gap distinct from the large pseudogap. *Phys. Rev. B* **84**, 014522, doi:10.1103/physrevb.84.014522 (2011).
50. Kim, E.-A. *et al.* Theory of the nodal nematic quantum phase transition in superconductors. *Phys. Rev. B* **77**, 184514, doi:10.1103/PhysRevB.77.184514 (2008).
51. Bourges, P. & Sidis, Y. Novel magnetic order in the pseudogap state of high- copper oxides superconductors. *Comptes Rendus Physique* **12**, 461–479, doi:10.1016/j.crhy.2011.04.006 (2011).
52. Sidis, Y. & Bourges, P. Evidence for Intra-Unit-Cell magnetic order in the pseudo-gap state of high-$T_c$ cuprates. *Journal of Physics: Conference Series* **449**, 012012, doi:10.1088/1742-6596/449/1/012012 (2013).
53. Chowdhury, D. & Sachdev, S. The Enigma of the Pseudogap Phase of the Cuprate Superconductors. In *50th Karpacz Winter School of Theoretical Physics* 1–43 (World Scientific, 2015).
54. Rice, T. M., Yang, K.-Y. & Zhang, F. C. A phenomenological theory of the anomalous pseudogap phase in underdoped cuprates. *Rep. Prog. Phys.* **75**, 016502, doi:10.1088/0034-4885/75/1/016502 (2012).
55. Abanov, A. & Chubukov, A. V. Spin-Fermion Model near the Quantum Critical Point: One-Loop Renormalization Group Results. *Phys. Rev. Lett.* **84**, 5608–5611, doi:10.1103/PhysRevLett.84.5608 (2000).
56. Hinton, J. P. *et al.* The rate of quasiparticle recombination probes the onset of coherence in cuprate superconductors. *Sci. Rep.* **6**, 23610, doi:10.1038/srep23610 (2016).
57. Montiel, X., Kloss, T. & Pépin, C. Effective SU(2) theory for the pseudo-gap state. *Phys. Rev. B* **95**, 104510, doi:10.1103/PhysRevB.95.104510 (2017).

### Acknowledgements
We thank Y. Sidis, H. Alloul and Ph. Bourges, for stimulating discussions. This work has received financial support from LabEx PALM (ANR-10-LABX-0039-PALM), ANR project UNESCOS ANR-14-CE05-0007, as well as the grant Ph743-12 of the COFECUB. The authors also like to thank the IIP (Natal, Brazil), the Aspen Center for Physics and the Perimeter Institute (Ontario, Canada), where parts of this work were done, for hospitality. This work is supported by the ERC, under grant agreement AdG-694651-CHAMPAGNE.

### Author Contributions
X.M., T.K. and C.P. contributed equally to this work. All authors have performed the calculations and analysed the results. All the authors have written and reviewed the manuscript.

### Additional Information
**Supplementary information** accompanies this paper at doi:10.1038/s41598-017-01538-1

**Competing Interests:** The authors declare that they have no competing interests.

**Publisher's note:** Springer Nature remains neutral with regard to jurisdictional claims in published maps and institutional affiliations.